\pdfoutput=1

\documentclass[twocolumn,aps,pra,showpacs,superscriptaddress]{revtex4}

\usepackage{graphicx}
\usepackage{amsmath,amssymb}
\usepackage[colorlinks=true,citecolor=blue,urlcolor=blue]{hyperref}

\def\endproof{\vrule height6pt width6pt depth0pt}

\begin{document}



\title{Kochen-Specker set with seven contexts}


\author{Petr Lison\v{e}k}
 \email{plisonek@sfu.ca}
 \affiliation{Department of Mathematics, Simon Fraser University, Burnaby, BC Canada V5A 1S6}

\author{Piotr Badzi{\c a}g}
 \affiliation{Department of Physics, Stockholm University, S-10691 Stockholm, Sweden}

\author{Jos\'{e} R. Portillo}
 \affiliation{Departamento de Matem\'{a}tica Aplicada I, Universidad de Sevilla, E-41012 Sevilla, Spain}

\author{Ad\'an Cabello}
 \email{adan@us.es}
 \affiliation{Departamento de F\'{\i}sica Aplicada II, Universidad de Sevilla, E-41012 Sevilla, Spain}


\date{\today}



\begin{abstract}
The Kochen-Specker (KS) theorem is a central result in quantum theory and has applications in quantum information. Its proof requires several yes-no tests that can be grouped in contexts or subsets of jointly measurable tests. Arguably, the best measure of simplicity of a KS set is the number of contexts. The smaller this number is, the smaller the number of experiments needed to reveal the conflict between quantum theory and noncontextual theories and to get a quantum vs classical outperformance. The original KS set had 132 contexts. Here we introduce a KS set with seven contexts and prove that this is the simplest KS set that admits a symmetric parity proof.
\end{abstract}



\maketitle


\section{Introduction}


The Kochen-Specker (KS) theorem \cite{Specker60,KS67} underlies a subtle but fundamental difference between classical and quantum theories. It shows that when describing systems with more than two distinguishable states, quantum theory, unlike its classical counterpart, is incompatible with the assumption of outcome noncontextuality. This means that there are quantum tests whose outcomes cannot be predefined prior to the actual tests in a way that they do not depend on the test's context, that is, on the choice of jointly measurable tests which might be performed together.

A standard proof of the theorem relies on a construction of a so-called KS set of quantum yes-no tests. The tests of a KS set are represented by rank-one projectors (or by the corresponding vectors), which are designed in a way making an assignment of the outcomes satisfying outcome noncontextuality impossible. More precisely, a {\em KS set} in dimension $d$ is defined as a set $S$ of $d$-dimensional complex vectors, with $d \ge 3$ and with the property that there is no map $f: S \rightarrow \{0,1\}$ such that, for any context (represented by an orthogonal basis in $S$), one and only one of the vectors is mapped to~1 \cite{PMMM05}.

Besides KS sets, there are two other ways of proving the KS theorem. One uses general operators instead of rank-one projectors \cite{Peres90,Mermin90}. Proofs of this type can be expressed in terms of KS sets \cite{Peres91,KP95}. The other way is based on sets of vectors that permit one to derive a noncontextuality (NC) inequality violated by any quantum state \cite{YO12,BBC12,KBLGC12}. These sets of vectors are either KS sets or subsets of them (see Ref.~\cite{Cabello11} for details). In other words, KS sets are behind all types of proofs of the KS theorem.

The construction of KS sets is highly relevant for the foundations of physics, not only because KS sets provide a proof by contradiction of the incompatibility between quantum theory and non-contextual realism (the KS theorem), but also because, assisted with maximally entangled states, KS sets provide a proof by contradiction of the incompatibility between quantum theory and local realism (the so-called KS theorem with locality \cite{HR83} or free-will theorem \cite{CK06,CK09}). KS sets can also be used to design experimental tests to show the quantum state-independent violation of NC inequalities \cite{Cabello08,BBCP09,KZGKGCBR09,ARBC09,DHANBSC13} and to design experimental tests for detecting fully nonlocal correlations \cite{AGACVMC12}. In quantum information, KS sets are used in quantum pseudotelepathy nonlocal games \cite{RW04,SS12}; in games with quantum state-independent advantage \cite{DHANBSC13}, for providing security against classical attacks to quantum cryptographic protocols based on complementarity \cite{Svozil10,CDNS11}; and for single-shot entanglement-assisted zero-error communication \cite{CLMW10,MSS13}.

The KS set in the original proof of the KS theorem \cite{KS67} contains $117$ vectors. This number is too high for a proof of such a fundamental result and also for practical applications. This motivated the search for more economical KS sets and simpler proofs of the KS theorem. Recently, it has been shown \cite{PMMM05,Cabello06,AOW11} that the KS set with the smallest number of vectors has 18 vectors in $d=4$. A set like that was introduced in Ref.~\cite{CEG96}. It has also been proven \cite{Cabello11} that the simplest proof of the KS theorem with a set of vectors which is not a KS set needs 13 vectors in $d=3$. A proof like that was introduced in Ref.~\cite{YO12}.

Nevertheless, it has been frequently pointed out \cite{PMMM05,Larsson02,Held09} that the above assessments of the proof's simplicity are not the most relevant, since the proofs of the KS theorem tacitly refer to many more vectors than those explicitly stated. This is so because the traditional way of counting vectors only takes into account those vectors that do not admit a KS valuation (in the case of KS sets) or that explicitly appear as variables in the state-independent NC inequality (in the case of proofs with sets that are not KS sets). As remarked in Ref.~\cite{Held09}, ``[t]his question of the actual size of a concrete KS set is important not so much for determining the record of the smallest such set, but for an experimental realisation, which actually involves procedures equivalent to basis.'' According to this observation, the physically relevant measure of simplicity of a proof of the KS theorem is the number of {\em bases}, which corresponds to the number of {\em contexts} in the KS set.

In this sense, the original proof of the KS theorem required 132 contexts in $d=3$ \cite{KS67}. The KS sets in $d=3$ with the smallest number of vectors require 40 \cite{Peres91}, 37 \cite{CK95}, and 36 contexts \cite{Bub97}, respectively. The 13-vector proof of Ref.~\cite{YO12} requires 16 contexts. The KS set with the smallest number of vectors requires nine contexts \cite{CEG96}. No other known proof requires a smaller number of contexts. Significantly, these last two sets are, so far, the only ones that have been implemented in experiments \cite{DHANBSC13,ZUZAWDSDK13}.

A fundamental open question is, {\em Which is the proof of the KS theorem with the smallest number of bases?}
The aim of this article is to answer this question. In Sec.~\ref{Sec2} we show that there is a KS set requiring only seven bases in $d=6$. In Sec.~\ref{Sec3}, we prove that, up to two natural assumptions, there is no KS set or proof of the KS theorem with yes-no tests requiring a smaller number of contexts. In addition, in Sec.~\ref{Sec4}, we use the KS set introduced in Sec.~\ref{Sec2} to derive a NC inequality which is violated by any quantum state in $d=6$.


\begin{figure}[bt]
\vspace{-3.9cm}
\centering
\centerline{\includegraphics[scale=0.44]{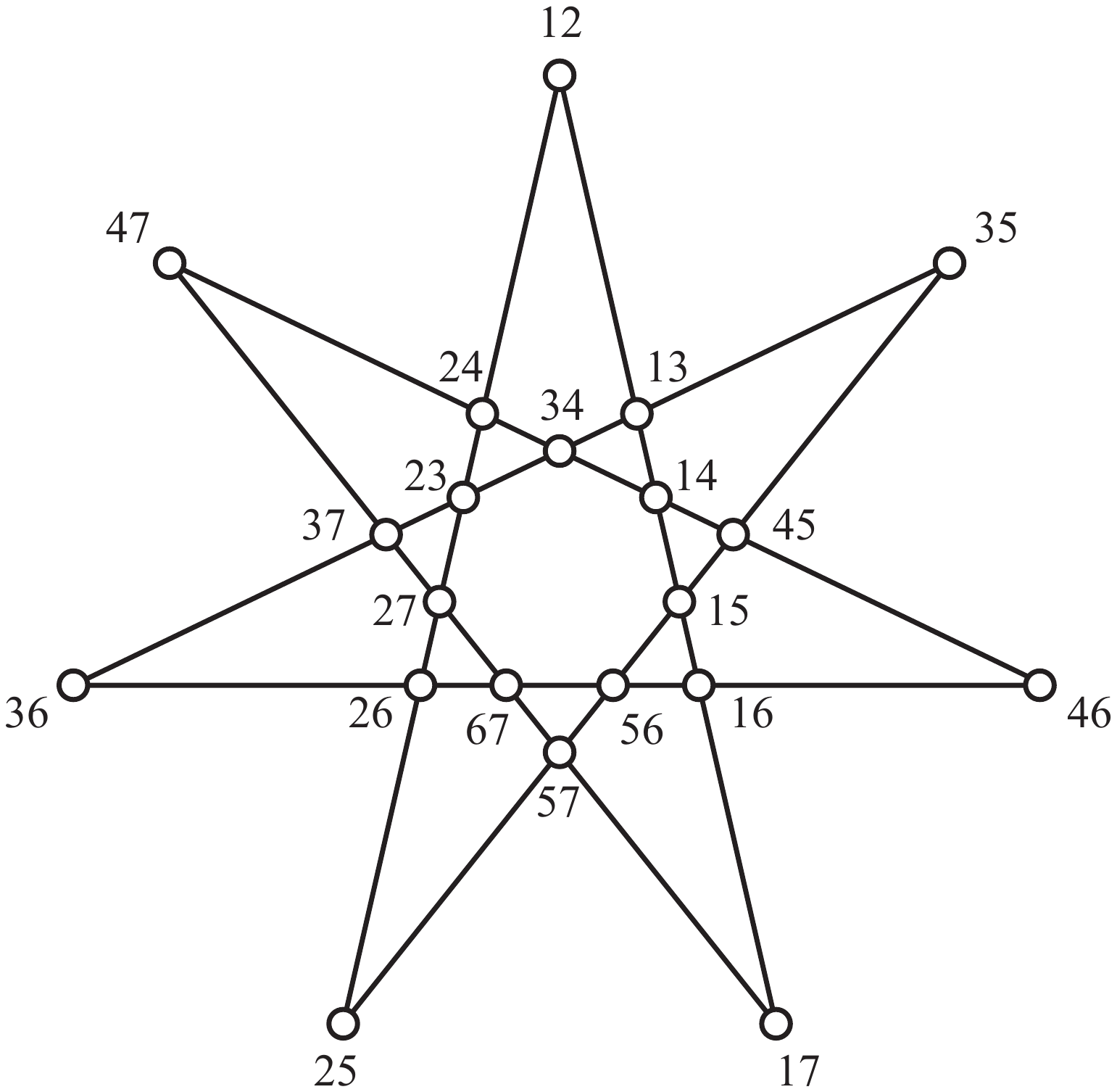}}
\vspace{-2.8cm}
\caption{\label{Fig1}Orthogonality relations between the vectors of the seven-context KS set. Vectors are represented by nodes, and contexts (bases) are represented by straight lines. The pair of numbers $ij$, with $i<j$, denotes the vector common to bases $B_i$ and $B_j$ in (\ref{Vi}).}
\end{figure}


\section{A 7-context KS set}
\label{Sec2}


Consider the following 7 orthogonal bases in $d=6$:
\begin{widetext}
\begin{subequations}
 \label{Vi}
\begin{align}
 B_1=&\left\{(1,0,0,0,0,0),(0,1,0,0,0,0),(0,0,1,0,0,0),(0,0,0,1,0,0),(0,0,0,0,1,0),(0,0,0,0,0,1)\right\},\\
 B_2=&\left\{(1,0,0,0,0,0),(0,0,1,1,1,1),(0,1,0,1,\omega,\omega^2),(0,1,1,0,\omega^2,\omega),(0,1,\omega,\omega^2,0,1),(0,1,\omega^2,\omega,1,0)\right\},\\
 B_3=&\left\{(0,1,0,0,0,0),(0,0,1,1,1,1),(1,0,0,1,\omega^2,\omega),(1,0,1,0,\omega,\omega^2),(1,0,\omega^2,\omega,0,1),(1,0,\omega,\omega^2,1,0)\right\},\\
 B_4=&\left\{(0,0,1,0,0,0),(0,1,0,1,\omega,\omega^2),(1,0,0,1,\omega^2,\omega),(1,1,0,0,1,1),(\omega,\omega^2,0,1,0,1),(\omega^2,\omega,0,1,1,0)\right\},\\
 B_5=&\left\{(0,0,0,1,0,0),(0,1,1,0,\omega^2,\omega),(1,0,1,0,\omega,\omega^2),(1,1,0,0,1,1),(\omega^2,\omega,1,0,0,1),(\omega,\omega^2,1,0,1,0)\right\},\\
 B_6=&\left\{(0,0,0,0,1,0),(0,1,\omega,\omega^2,0,1),(1,0,\omega^2,\omega,0,1),(\omega,\omega^2,0,1,0,1),(\omega^2,\omega,1,0,0,1),(1,1,1,1,0,0)\right\},\\
 B_7=&\left\{(0,0,0,0,0,1),(0,1,\omega^2,\omega,1,0),(1,0,\omega,\omega^2,1,0),(\omega^2,\omega,0,1,1,0),(\omega,\omega^2,1,0,1,0),(1,1,1,1,0,0)\right\},
\end{align}
\end{subequations}
\end{widetext}
where $\omega=e^{2 \pi i/3}$. For simplicity, normalization factors are omitted.

The seven bases in (\ref{Vi}) contain 21 different vectors. Each vector belongs to two bases. The proof that the vectors in (\ref{Vi}) constitute a KS set is straightforward: to map one and only one of the vectors in each basis to $1$, only seven vectors in (\ref{Vi}) must be mapped to $1$. However, since each vector belongs to two bases, any mapping forces one to map to $1$ an {\em even} number of vectors. This makes the mapping impossible.

The new KS set is represented in Fig.~\ref{Fig1}. The seven bases are represented by straight lines and each vector is represented as a node. The pair of numbers $ij$, with $i<j$, denotes the vector common to bases $B_i$ and $B_j$.

This seven-context 21-vector KS set does not only improve the current record of contexts in any $d$ \cite{CEG96} and the current record of vectors in $d=6$ \cite{CEG05} but, more importantly, as proven in the next section, constitutes the symmetric parity proof of the KS theorem (defined later) requiring the smallest number of contexts in any $d$.


\section{Proof that the 7-context KS set is the symmetric parity proof with the smallest number of contexts}
\label{Sec3}


Any set of $d$-dimensional vectors can be associated to a graph $G$ in which vectors are represented by vertices such that two vertices in $G$ are adjacent if and only if the vectors that they represent are orthogonal. Orthogonal bases then correspond to cliques of size $d$ in $G$ (i.e., sets of $d$ mutually adjacent vertices). A set of $d$-dimensional vectors allows for a proof of the KS theorem only if, in the corresponding $G$, the minimum number of colors needed to color all vertices avoiding adjacent vertices to have the same color [i.e., the chromatic number of $G$, $\chi(G)$] is strictly greater than $d$ \cite{Cabello11}.

The KS set with the smallest number of bases previously known, namely the one in Ref.~\cite{CEG96}, has two extra properties which make it particularly appealing: (1) Its $G$ is vertex transitive, i.e., given any two vertices $v_1$ and $v_2$ of the vertex set of $G$, denoted as $V$, there is some automorphism $f:V \rightarrow V$ such that $f(v_1) = v_2$; and (2) its $G$ has an odd number of cliques of size $d$, while each vertex belongs exactly to an even number of them. Any proof of the KS theorem having property (2) is called a {\em parity proof,} since the proof immediately follows from a simple parity argument. Any proof of the KS theorem having properties (1) and (2) is called a {\em symmetric parity proof.} Any KS set having properties (1) and (2) is called a {\em symmetric parity KS set.}

The first parity KS set was found by Kernaghan \cite{Kernaghan94} and the first symmetric parity KS set is the one in Ref.~\cite{CEG96}. Parity KS sets for systems of two, three, and four qubits have received special attention \cite{WA10,WAMP11,WA11a,WA11b,Planat12}.

Our purpose is to prove that the KS set presented in the previous section is the symmetric parity proof of the KS theorem with the smallest number of contexts.

The proof is as follows. For any parity KS set, the graph $G$ in which adjacent vertices represent orthogonal vectors is a fully contextual graph \cite{ADLPBC12}, namely, $\alpha(G)<\vartheta(G)=\alpha^*(G)$, where $\alpha(G)$, $\vartheta(G)$, and $\alpha^*(G)$ are the independence number, Lov\'asz number, and fractional packing number of $G$, respectively (for definitions, see Appendix~\ref{Appendix1} and Refs.~\cite{CSW10,CSW14}). We generate {\em all} connected graphs with at most 31 vertices that are both fully contextual and vertex transitive (see Appendix~\ref{Appendix2} for details). The number of graphs that are both fully contextual and vertex transitive are in column ``FCVT'' in Table~\ref{Table1}.


\begin{table}[tb]
\caption{\label{Table1}``FCVT'' indicates the number of fully contextual vertex-transitive connected graphs on $n\le31$ vertices. If a particular $n$ is not in the table, this means that there are no FCVT graphs with $n$ vertices.
``PFCVT'' gives the number of FCVT graphs with an odd number of cliques of maximum size and every vertex belonging to an even number of cliques of maximum size. The number of cliques of maximum size is indicated in brackets. ``KS sets'' indicates the number of symmetric parity KS sets. The dimension of the vectors is in brackets. Numbers in boldface correspond to graphs analyzed in detail. ``?'' indicates that we have not analyzed these graphs, since they cannot produce simpler proofs of the KS theorem.}
\begin{ruledtabular}
\begin{tabular}{cccc}
Vertices & FCVT & PFCVT (bases) & KS sets (dim.)\\
 \hline
10 & 1 & {\bf 1 (5)} & {\bf 0} \\
16 & 2 & 0 & 0\\
18 & 3 & {\bf 1 (9)} & {\bf 1 (4)}\\
20 & 24 & {\bf 5 (5)} & {\bf 0}\\
   &    & 1 (25) & 0\\
21 & 4 & {\bf 3 (7)} & {\bf 1 (6)}\\
24 & 113 & 0 & 0\\
25 & 5 & 0 & 0\\
26 & 11 & 7 (13) & ? (4)\\
27 & 22 & 12 (9) & ? (6)\\
   &    & 1 (27) & ? (6)\\
28 & 46 & 0 & 0\\
30 & 468 & 30 (15) & ? (4)\\
   &     & 3 (45) & 0 (4)\\
   &     & 1 (405) & 0 (4)\\
   &     & {\bf 14 (5)} & {\bf 0} \\
   &     & 1 (125) & 0 (12) \\
\end{tabular}
\end{ruledtabular}
\end{table}


If we add the restriction that the number of cliques of maximum size is odd and that every vertex belongs to an even number of them, then the number of graphs reduces substantially (see column ``PFCVT'' in Table~\ref{Table1}). Only a few of them have nine or less bases.

The first interesting graph, in boldface in Table~\ref{Table1}, is a ten-vertex graph called the Johnson $J(5,2)$ graph. It does not correspond to a KS set, since the maximum size of the cliques is 4, while the graph does not admit a representation with vectors of $d=4$, but requires vectors of dimension 6 \cite{Cabello13} (see Appendix~\ref{Appendix3} for a proof). This graph also cannot be used for a proof of the KS theorem without KS sets, since its chromatic number is~5.

The second interesting graph, in boldface in Table~\ref{Table1}, corresponds to the nine-basis 18-vector KS set in Ref.~\cite{CEG96}.

Then we have five graphs corresponding to five-basis 20-vector sets. None of them corresponds to a parity KS set, since all of them contain a graph that does not admit a representation in $d=8$. In addition, none of them can be used for a parity proof without KS sets, since the chromatic number of the common graph is~10, while there is no representation of it in $d=9$ (see Appendix~\ref{Appendix3} for a proof).

The fourth interesting entry in Table~\ref{Table1} are the three graphs corresponding to seven-basis 21-vector sets. All these graphs have the graph corresponding to the seven-basis 21-vector KS set introduced in this article as a subgraph.

For graphs with a higher number of vertices, we have focused on those graphs that have exactly five cliques of maximum size in column ``PFCVT'' in Table~\ref{Table1}. There are 14 of them. All of them correspond to five-basis 30-vertex sets. However, none of them is a parity KS set, since all of them contain a subgraph that does not admit a representation in $d=12$. In addition, none of them can be used for a parity proof without KS sets, since the common subgraph has chromatic number 15, while there is no representation in $d=14$ (see Appendix~\ref{Appendix3} for a proof).

Our exploration is exhaustive up to graphs on 31 vertices. Moreover, for symmetric fully contextual graphs with more vertices, the only PFCVT graphs with exactly five cliques of maximum size are $10 k$-vertex graphs with five cliques of maximum size $4k$ containing an orthogonality structure that can be represented by a Johnson $J(5,2)$ graph, assuming that each vertex of the Johnson represents a clique of size $k$. The only other vertex-transitive graphs with exactly five cliques of maximum size are the $5k$-vertex graphs with five cliques of maximum size $2k$ that can be represented by a pentagon, assuming that each vertex in the pentagon represents a clique of size $k$. However, these graphs are not fully contextual.

The $10 k$-vertex $J(5,2)$ graphs have already appeared in Table~\ref{Table1} for $k=1,2,3$. None of these graphs can be a KS set, since they do not admit a representation in dimension $4 k$ (see Appendix~\ref{Appendix3} for a proof). Moreover, none of them can be a proof of the KS theorem without KS sets, since these graphs have chromatic number $5k$, but do not admit a representation in dimension $5 k-1$ (see Appendix~\ref{Appendix3} for a proof).

Clearly, no symmetric parity proof exists with exactly three cliques: Corollary 7.5.2 in Ref.~\cite{GR01} implies that for a $n$-vertex, vertex-transitive graph $G$ we have $\alpha(G)\omega(G)\leq n$. As always, $\alpha(G)$ denotes the size of the largest independent set in $G$ (the independence number of $G$) and $\omega(G)$ denotes the clique number of $G$, i.e., the size of the largest clique in $G$. Since KS graphs are not complete, we have $\alpha(G) \geq 2$, and $\omega(G)\leq n/2$ follows. Assuming that each vertex is in two or more cliques leads to $2 n \leq 3 \omega(G) \leq 3 n/2$, which is impossible.

From this, we can conclude that the KS set presented in this article is the symmetric parity KS set and the symmetric parity proof of the KS theorem having the smallest number of bases.


\section{Quantum state-independent noncontextuality inequality}
\label{Sec4}


Here we obtain a quantum state-independent NC inequality starting from the KS set introduced before. There is a general method for producing a quantum state-independent NC inequality from any KS set \cite{BBCP09}. However, here we exploit the extra symmetries of the seven-context KS set to end up with a very compact inequality.

Consider 21 observables $A_{ij}$, with $i,j=1,\ldots,7$ and $i<j$, each with possible results $-1$ and $+1$. For any theory satisfying outcome noncontextuality, the following NC inequality is satisfied:
\begin{equation}
\begin{split}
 S=&-\langle A_{12} A_{13} A_{14} A_{15} A_{16} A_{17} \rangle
-\langle A_{12} A_{23} A_{24} A_{25} A_{26} A_{27} \rangle \\
 &-\langle A_{13} A_{23} A_{34} A_{35} A_{36} A_{37} \rangle
 -\langle A_{14} A_{24} A_{34} A_{45} A_{46} A_{47} \rangle \\
 &-\langle A_{15} A_{25} A_{35} A_{45} A_{56} A_{57} \rangle
 -\langle A_{16} A_{26} A_{36} A_{46} A_{56} A_{67} \rangle \\
 &-\langle A_{17} A_{27} A_{37} A_{47} A_{57} A_{67} \rangle \stackrel{\mbox{\tiny{ NCHV}}}{\leq} 5,
 \label{NCI}
\end{split}
\end{equation}
where $\langle \ldots \rangle$ denotes the mean value of the product of the outcomes.

By choosing the following quantum observables,
\begin{equation}
 A_{ij}=2 |v_{ij}\rangle\langle v_{ij}|-\openone, \label{observable}
\end{equation}
with the normalized version of the vectors $|v_{ij}\rangle$ in (\ref{Vi}), we obtain that, in quantum theory, for any quantum state in $d=6$,
\begin{equation}
 S \stackrel{\mbox{\tiny{Q}}}{=} 7.
\end{equation}
The quantum violation of inequality (\ref{NCI}) can be tested with four sequential measurements on a six-dimensional quantum system. A simpler experiment to test this KS set consists of implementing the corresponding game with state-independent quantum advantage \cite{DHANBSC13}.

The 21-vertex graph in Fig.~\ref{Fig1} contains 21 Johnson $J(5,2)$ graphs induced. This can be seen by removing from the graph in Fig.~\ref{Fig1} all the nodes in any two straight lines (i.e., all the vectors of any pair of orthogonal bases).
Therefore, inequality (\ref{NCI}) can be considered a state-independent version of the twin inequality introduced in Ref.~\cite{Cabello13}. From a different perspective, the ten-question set in \cite{Cabello13} can be considered a five-context state-dependent KS set (as defined in Ref.~\cite{CG96}) that is a subset of the seven-context state-independent KS set introduced here. The 21-vertex graph in Fig.~\ref{Fig1} was also considered in Ref.~\cite{AFLS12} without noticing that it can represent a KS set.


\section{Conclusions}
\label{Sec5}


Arguably, the best measure of simplicity of a KS set is the number of contexts. In this article we have presented a KS set with the smallest number of contexts known: seven. In addition, we have proven that our KS set is not only the simplest KS set that admits a symmetric parity proof, but also the simplest symmetric set of yes-no tests that can be used to prove the KS theorem. Finally, we have used our KS set to derive a compact NC inequality violated by any quantum state in dimension $6$.

We think that the KS set introduced in this article is important for foundations of quantum theory and may have applications in quantum information processing. It is surprising that it has remained undiscovered for so long.


\begin{acknowledgments}
 We thank J.-{\AA}. Larsson, M. Terra Cunha, A. Sol\'{\i}s, and M. Waegell for useful conversations. This work was supported by the Natural Sciences and Engineering Research Council of Canada (NSERC), the Collaborative Research Group ``Mathematics of Quantum Information'' of the Pacific Institute for the Mathematical Sciences (PIMS), the project No.\ FIS2011-29400 (Ministerio de Econom\'{\i}a y Competitividad, Spain) with Fonds Europ\'een de D\'eveloppment \'Economique et R\'egional, and the FQXi large grant project ``The Nature of Information in Sequential Quantum Measurements''.
\end{acknowledgments}


\appendix


\section{Parity KS sets are represented by fully contextual graphs}
\label{Appendix1}


The {\em independence number} of a graph $G$, denoted as $\alpha(G)$, is the maximum number of nonadjacent vertices in $G$.

The {\em Lov\'asz number} of a graph $G$ with vertex set $V$, denoted as $\vartheta(G)$, is defined as $\vartheta(G):= \max \sum_{i \in V} | \langle \psi | v_i \rangle |^2$, where the maximum is taken over all sets of unit vectors $|v_i \rangle \in \mathbb{R}^d$ such that $\langle v_i | v_j \rangle=0$ for all pairs $i,j$ of adjacent vertices in $V$, all unit vectors $|\psi \rangle \in \mathbb{R}^d$, and all $d$.

The {\em fractional packing number} of a graph $G$, denoted as $\alpha^*(G)$, is defined as $\alpha^*(G):=\max \sum_{i \in V} p_i$, where the maximum is taken over all $p_i \geq 0$ and for all cliques $C$ of $G$, under the restriction $\sum_{i \in C} p_i \leq 1$.

A graph $G$ is {\em fully contextual} if $\alpha(G)<\vartheta(G)=\alpha^*(G)$ \cite{ADLPBC12}.

{\em Lemma:} The orthogonality graph $G$ of a symmetric parity KS set is fully contextual.

{\em Proof:} If $G$ corresponds to an $n$-vector KS set in dimension $d$, then $\vartheta(G)= n/d$. This follows from the fact that $\vartheta(G)$ equals the quantum maximum of the sum $S=\sum_{i \in V} P_{\rho}(|u_i \rangle \langle u_i|=1)$ of probabilities of obtaining outcome 1 when rank-one projectors $|u_i \rangle \langle u_i|$, with unit vectors $|u_i \rangle \in \mathbb{C}^d$ such that $\langle u_i | u_j \rangle=0$ for all pairs $i,j$ of adjacent vertices in $V$, are measured on a physical system prepared in a quantum state $\rho$. For a KS set, the value of $S$ is the same for any quantum state $\rho$. $S$ is $n/d$ for a maximally mixed state $\rho=\openone/d$, where $\openone$ is the $d \times d$ identity matrix.

If $G$ corresponds to a KS set, then $\alpha(G) < \vartheta(G)$. $\alpha(G)$ is the maximum number of vectors in the KS set that can be mapped to 1 so that no two orthogonal vectors are both mapped to 1. For a KS set, this number must be strictly smaller than $S$.

If $G$ is vertex transitive, then $\alpha^*(G)=n/d$.\hfill \endproof


\section{How we made Table~\ref{Table1}}
\label{Appendix2}


For making Table~\ref{Table1} we used a previously existing database of vertex-transitive graphs \cite{database}. All the results in this database have been checked by independent authors, except for the graphs with 27, 28, and 30 vertices. On these graphs, we calculated $\alpha(G)$, $\vartheta(G)$, $\alpha^*(G)$, and $\chi(G)$. For calculating $\vartheta(G)$ we used {\tt DSDP} \cite{DSDP}. For calculating $\alpha^*(G)$ we used the fact that, for a vertex-transitive graph on $n$ vertices, $\alpha^*(G)= n/\omega(G)$, where $\omega(G)$ is the clique number of $G$; for calculating $\omega(G)$, $\alpha(G)$, and $\chi(G)$ we used {\tt nauty} \cite{McKay90} and {\tt very\_nauty} \cite{verynauty}. Finally, we used {\tt Mathematica} \cite{Mathematica} for counting the cliques in the graphs of Table~\ref{Table1}.


\section{Proofs that some graphs in Table~\ref{Table1} cannot correspond to parity KS proofs}
\label{Appendix3}


The 20-vertex graph with five cliques of size 8 common to the five graphs on 20 vertices (and the 30-vertex subgraph with five cliques of size 12 common to the 14 graphs on 30 vertices) indicated in boldface in Table~\ref{Table1} can be represented by the Johnson $J(5,2)$ graph assuming that each vertex of the Johnson actually represents a clique of size 2 (3). Similarly, the $10 k$-vertex graphs with five cliques of size $4k$, with $k=1,2,3,4,\ldots$, can be represented by the Johnson $J(5,2)$ graph assuming that each vertex of the Johnson actually represents a clique of size $k$.

However, there is no set of five bases in dimension $4 k$ which allows for this structure of orthogonality. To prove it, let us assign a $4k$-dimensional basis ${\cal B}_j$, with $j=0,\ldots, 4$ to each clique of size $4k$. Let the columns of matrix ${\cal B}_j$ represent the vectors of basis ${\cal B}_j$ and let ${\cal B}_0$ be the coordinate basis in $\mathbb{C}^{4k}$. Then, matrix ${\cal B}_0$ can be chosen as the $4k \times 4k$ identity. With this fixed, the structure of the graph requires that the remaining matrices have the following block structures:
\begin{equation}
\begin{split}
 &{\cal B}_1 = \left(
  \begin{array}{cccc}
    \openone & 0 & 0 & 0 \\
    0 & 0 & C_1 & C_2 \\
    0 & A_1 & 0 & A_2 \\
    0 & B_1 & B_2 & 0 \\
  \end{array}
\right),\;\;\; {\cal B}_2 = \left(
  \begin{array}{cccc}
    0 & 0 & D_1 & D_2 \\
    0 & \openone & 0 & 0 \\
    A_1 & 0 & 0 & A_3 \\
    B_1 & 0 & B_3 & 0 \\
  \end{array}
\right), \\
 &{\cal B}_3 = \left(
  \begin{array}{cccc}
    0 & D_1 & 0 & D_3 \\
    C_1 & 0 & 0 & C_3 \\
    0 & 0 & \openone & 0 \\
    B_2 & B_3 & 0 & 0 \\
  \end{array}
\right),\;\;\; {\cal B}_4 = \left(
  \begin{array}{cccc}
    0 & D_2 & D_3 & 0 \\
    C_2 & 0 & C_3 & 0 \\
    A_2 & A_3 & 0 & 0 \\
    0 & 0 & 0 & \openone \\
  \end{array}
\right).
 \label{unoq}
\end{split}
\end{equation}
The entries represent $k\times k$ matrices, and $\openone$ is the $k\times k$ identity. Due to orthogonality of the columns in each of the matrices, the nonzero columns in the $k\times k$ blocks denoted by the same letter must be orthogonal. Three blocks contain together $3k$ columns, but the columns in each of the blocks must be orthogonal to the columns in the other two blocks. Thus the number of linearly independent columns in the three blocks cannot exceed the number of rows in the blocks, $k$, i.e.,
\begin{subequations}
\label{condition1}
\begin{align}
  d_1(D_1) + d_1(D_2) + d_1(D_3) & \leq k, \\
  d_2(C_1) + d_2(C_2) + d_2(D_3) & \leq k, \\
  d_3(A_1) + d_3(C_2) + d_3(D_2) & \leq k, \\
  d_4(A_1) + d_4(C_1) + d_4(D_1) & \leq k,
\end{align}
\end{subequations}
where $d_i(X_j)$ denotes the column rank corresponding to block row $i$ and the top nonzero block entry $X_j$. By summing up inequalities (\ref{condition1}) and rearranging the terms one obtains
\begin{equation}
\label{condition1Sum}
\begin{split}
  & d_1(D_1) + d_4(D_1) + d_1(D_2) + d_3(D_2)+ d_1(D_3) \\
  & + d_2(D_3) + d_2(C_1) + d_4(C_1) + d_2(C_2) + d_3(C_2) \\
  & + d_3(A_1) + d_4(A_1) \leq 4k.
\end{split}
\end{equation}

On the other hand, the number of independent columns in each column block, $d(X_j)$, is equal to $k$. This number is upper bounded by the number of independent subcolumns corresponding to this block. Thus $d(D_1) \leq d_1(D_1) + d_4(D_1)$, and the same is true for the other columns. The left-hand side of inequality (\ref{condition1Sum}) is therefore not less than $6k$, which leads to $6k \leq 4k$, a contradiction for all $k \geq 1$.

To prove that there is no set of five bases in dimension $5 k-1$ that allows for this structure of orthogonality, notice that if the structure is embedded in a larger Hilbert space than $\mathbb{C}^{4k}$, say $\mathbb{C}^{4k+p}$, then one needs to augment matrix ${\cal B}_0$ with $p$ rows of zeros and the remaining matrices ${\cal B}_j$ with $p$ rows according to the following block structure
\begin{equation}
\begin{split}
 &{\cal B}_1 = \left(
  \begin{array}{cccc}
    \openone & 0 & 0 & 0 \\
    0 & 0 & C_1 & C_2 \\
    0 & A_1 & 0 & A_2 \\
    0 & B_1 & B_2 & 0 \\
    0 & X & Y & Z \\
  \end{array}
\right),\;\;\; {\cal B}_2 = \left(
  \begin{array}{cccc}
    0 & 0 & D_1 & D_2 \\
    0 & \openone & 0 & 0 \\
    A_1 & 0 & 0 & A_3 \\
    B_1 & 0 & B_3 & 0 \\
    X & 0 & T & S \\
  \end{array}
\right), \\
 &{\cal B}_3 = \left(
  \begin{array}{cccc}
    0 & D_1 & 0 & D_3 \\
    C_1 & 0 & 0 & C_3 \\
    0 & 0 & \openone & 0 \\
    B_2 & B_3 & 0 & 0 \\
    Y & T & 0 & U \\
  \end{array}
\right),\;\;\; {\cal B}_4 = \left(
  \begin{array}{cccc}
    0 & D_2 & D_3 & 0 \\
    C_2 & 0 & C_3 & 0 \\
    A_2 & A_3 & 0 & 0 \\
    0 & 0 & 0 & \openone \\
    Z & S & U & 0 \\
  \end{array}
\right).
 \label{unoqp}
\end{split}
\end{equation}
The additional blocks contain $p$ rows and $k$ columns. Due to the new rows, inequalities (\ref{condition1}) nor read
\begin{subequations}
\label{condition2}
\begin{align}
 d_{15}(D_1) + d_{15}(D_2) + d_{15}(D_3) & \leq k + p, \\
 d_{25}(C_1) + d_{25}(C_2) + d_{25}(D_3) & \leq k + p, \\
 d_{35}(A_1) + d_{35}(C_2) + d_{35}(D_2) & \leq k + p, \\
 d_{45}(A_1) + d_{45}(C_1) + d_{45}(D_1) & \leq k + p,
\end{align}
\end{subequations}
with $d_{ij}(X)$ denoting the number of independent columns in block rows $i$ and $j$ of block column $X$. When summing both sides of (\ref{condition2}) one should notice that the contribution from the additional rows appears twice in each block column. With this observation and the same reasoning as before, one now gets $6k \leq 4k+2p$, which for positive $k$ implies $p \geq k$.



\end{document}